\documentclass[prb,twocolumn,notitlepage,superscriptaddress,longbibliography]{revtex4-2}
\usepackage{amsmath}
\usepackage{amssymb}

\usepackage[unicode=true,colorlinks=true,citecolor=blue,urlcolor=blue]{hyperref}

\usepackage{bm}
\usepackage{epsfig}
 
\usepackage[normalem]{ulem}

\usepackage{amssymb}
\usepackage{hyperref}
\usepackage{color}
\usepackage{dcolumn,amsmath,amsthm,amscd,amsfonts,amssymb,epsfig,graphics,graphicx,eucal}

\newcommand{\rmi}{{\rm i}}
\newcommand{\rmd}{{\rm d}}
\newcommand {\e}{{\rm e}}

\begin{document}

\title{Optomechanical lasing and domain walls driven by exciton-phonon interactions}
\author{Alexey V. Yulin}
\affiliation{ITMO University 197101, Kronverksky pr. 49, St. Petersburg, Russian Federation}

\author{Alexander V. Poshakinskiy}
\affiliation{Ioffe Institute, St. Petersburg 194021, Russia}

\author{Alexander N. Poddubny}
\affiliation{Ioffe Institute, St. Petersburg 194021, Russia}
\affiliation{ITMO University 197101, Kronverksky pr. 49, St. Petersburg, Russian Federation}
\email{poddubny@coherent.ioffe.ru}

\date{\today}

\begin{abstract}
We study theoretically interaction of 
optically-pumped  excitons with acoustic waves in planar semiconductor nanostructures in the strongly nonlinear regime. We start with the multimode optomechanical lasing  regime for optical pump frequency {above} the exciton resonance and demonstrate broadband chaotic-like lasing spectra.
We also predict formation of propagating optomechanical domain walls driven by optomechanical nonlinearity for the optical pump  {below} the exciton resonance. 
Stability conditions for the domain walls  are examined analytically and are in agreement with direct numerical simulations. Our results apply  to nonlinear  sound propagation in the arrays of quantum wells or in the plane of Bragg semiconductor microcavities hosting excitonic polaritons.
\end{abstract}

\maketitle

\section{Introduction}
Semiconductor optomechanics with excitonic polaritons, hybrid half-light half-matter quasiparticles, exhibits now a rapid progress driven by the success of planar nanofabrication technology~\cite{Delsing_2019}. Polariton-driven phonon lasing~ \cite{Chafatinos2020} and dynamically tuned arrays of polariton parametric oscillators~\cite{Kuznetsov2020} have been recently demonstrated. The special feature of the polaritonic platform is the resonant photoelastic interaction mechanism, that is present in addition to the usual geometric one~\cite{Baker:14}:  both excitonic and optical component can interact with the acoustic wave, which can enhance the overall optomechanical coupling~\cite{Jusserand2015}. The optomechanical polaritonic systems have been predicted to amplify sound in a parity-time-symmetric fashion~\cite{Poshakinskiy2016} and should feature strong acoustic nonreciprocity~\cite{Poshakinskiy2017}.
However, most studies and observations of the nonlinear optomechanical dynamics for polaritons have been so far been performed in the  cavity optomechanics regime, when just several localized excitonic, photonic and acoustic modes interact with each other, see Ref.~\cite{vyatkin2021optomechanical} and references therein. 
Given recent tremendous progress in understanding of collective nonlinear optomechanical effects in arrays of coupled cavities~\cite{Marquardt2011,Danatan2012,Martens2013,Bagheri2013,Eggleton2013,Gan:16}, 
including synchronization phenomena~\cite{Bagheri2013,Roukes2014}, formation of solitons~\cite{Gan:16} and chimera states~\cite{Martens2013},
it is quite instructive to examine distributed polaritonic systems, where  acoustic waves  interact with excitons and light.

Here, we study theoretically the nonlinear optomechanical interaction between optically pumped excitons and propagating acoustic waves.
We assume that excitons have large mass and thus are quasi-localized in space. Such situation is typically realized in periodic semiconductor superlattices where the excitons are confined inside 
the quantum wells while the longitudinal acoustic wave can propagate freely along the structure normal~\cite{Jusserand2015}. Alternatively, our results can be applied to laterally propagating acoustic waves in the planar Bragg semiconductor microcavities~\cite{Kuznetsov2020}. We demonstrate, that the nonlinear optomechanical dynamics can be quite intricate depending on the pump frequency, length of the structure and boundary conditions. In addition to the well-known optomechanical lasing regime~\cite{Kippenberg2014} we predict formation of propagating subsonic optomechanical domain walls.

\section{Model and linear stability analysis}\label{sec:model}

We consider propagation of interacting light, excitons and acoustic waves along the normal $z$ of a planar periodic semiconductor nanostructure, such as an array of quantum wells. When the optical and acoustic wave lengths are greater than the structure period along the growth axis, the system can be considered as a continuous medium. The Lagrangian density for such system has the form 
\begin{align}\label{eq:L}
L &= \frac{1}{8\pi} \left[\frac{(\partial_t A)^2}{c^2} - (\partial_z A)^2\right] - \frac1c\, P\,\partial_t A \nonumber\\
&+ \frac{\rho}2 (\partial_t u)^2- \frac{E }2 (\partial_z u)^2 \nonumber\\
&+ \frac{f}{2} [(\partial_t P)^2 - (\omega_x + \Xi\, \partial_z u)^2P^2] \,.
\end{align}
The first term of the Lagrangian~\eqref{eq:L} corresponds to the free electromagnetic field described by vector potential $A(z,t)$, while the second term represents its interaction with exciton polarization $P(z,t)$. The second line of the Lagrangian~\eqref{eq:L} governs the field of mechanical displacement $u(z,t)$, with $\rho$ and $E$ being the medium density and Young's modulus, respectively.
The last line of Eq.~\eqref{eq:L} is the Lagrangian of the harmonic oscillator describing the exciton polarization. The oscillator frequency shifts linearly with the deformation $\partial_z u$; the deformation potential constant is $\Xi \approx 10\,\text{eV}$.  
We suppose that excitons cannot move along $z$ direction being strongly confined inside individual quantum wells. The normalization constant $f$ is related to longitudinal-transverse splitting $\omega_\text{LT}$ ~\cite{Ivchenko2005} as $f=2\pi/(\omega_x\omega_\text{LT})$. 

From the Lagrangian~\eqref{eq:L} the equation for the evolution of the fields $A(z,t)$, $u(z,t)$, and $P(z,t)$ are derived:
\begin{align}
& \partial_t^2 A - c^2 \partial_z^2 A = 4\pi c \partial_t P \,,\label{eq:A}\\
&  \partial_t^2 u -  s^2 \partial_z^2 u = (f\omega_x \Xi/\rho)\, \partial_z P^2  \,,\label{eq:u}\\
& \partial_t^2 P + \omega_x^2P = - 2\omega_x \Xi P \, \partial_z u -(1/fc)\, \partial_t A \,,\label{eq:P}
\end{align}
where $s = \sqrt{E/\rho}$ is the (longitudinal) sound velocity and we neglected the term quadratic in $\Xi$. 

The exciton frequency being much higher than acoustic one, we suppose $P(z,t) = P_1 (z,t) \e^{-\rmi\omega_x t} + \text{c.c.}$ and $A(z,t) = A_1 (z,t) \e^{-\rmi\omega_x t} + \text{c.c.}$ Then, exploiting the rotating wave approximation, we obtain from Eqs.~\eqref{eq:u}-\eqref{eq:P}
\begin{align}
&  \partial_t^2 u -  s^2 \partial_z^2 u = \frac{4\pi \Xi}{\omega_\text{LT}\rho}\, \partial_z |P_1|^2 \,, \label{eq:u1}\\
&\rmi\partial_t P_1  =  \Xi P_1 \, \partial_z u - \frac{\omega_\text{LT}}{4\pi} E_1 \,,\label{eq:P1} 
\end{align}
where $E_1(z,t) = (\rmi\omega_x/c) A_1(z,t)$ is the electric field. 
Instead of the displacement field $u(z,t)$ it is convenient to use a combined field $\xi(z,t) = \Xi \partial_z u + \epsilon |P_1|^2$, where $ \epsilon=4\pi \Xi^2/(\omega_\text{LT} E)$. That field is proportional to stress and accounts for both mechanic and excitonic contributions. Then, we obtain the equation set
\begin{align}
&  \partial_t^2 \xi +2\Gamma_s \partial_t \xi -  s^2 \partial_z^2 \xi = \epsilon \partial_t^2 |P_1|^2  \,, \label{eq:u2}\\
&\rmi\partial_t P_1  =   - \rmi \Gamma_x P_1 +\xi P_1 + \alpha |P_1|^2 P_1+G\,,\label{eq:P2} 
\end{align}
where $G(z,t) = (\omega_\text{LT}/4\pi) E_1(z,t)$ and we additionally introduced decay rates for exciton and sound, $\Gamma_x$ and $\Gamma_s$, respectively. It follows from Eq.~\eqref{eq:P2} that interaction with sound induces exciton non-linearity with $\alpha = - \epsilon$~\cite{Vishnevsky2011}. However, this kind of non-linearity for realistic parameters is suppressed by exciton-exciton repulsive interaction leading to the overall  $\alpha>0$. The effect of such non-linearity has been extensively studied, in particular in Refs.~\cite{Vishnevsky2011,Bobrovska2017,Yulin2019}.
In this paper, we are interested in the effect of exciton-sound interaction. Therefore, we omit the term proportional $\alpha$ in our calculations. We also  suppose that the electric field $E_1(z,t)$ is determined by the pump laser and neglect the back-action of the excitons, i.e., disregard Eq.~\eqref{eq:A}, which is justified provided that $\omega_\text{LT} < \Gamma_x$.

We consider the case of homogeneous excitation with frequency $\omega_p$, $E_1(z,t) = E \e^{-\rmi(\omega_p-\omega_x)t}$. First, we focus on the solution with spatially uniform stress $\xi(z,t) = \xi_0$ and exciton polarization $P_1(z,t) = b_0  \e^{-\rmi(\omega_p-\omega_x)t}$, where
\begin{align}
b_0 = \frac{G}{\omega_p-\omega_x - \xi_0 + \rmi\Gamma_x}  \,.
\end{align}
To analyze its stability, we investigate the dynamics of small corrections $\zeta$ and $a=a_1+\rmi a_2$ for the deformation field $\xi$ and the exciton field $P_1$. Their evolution is governed by the equations  
\begin{align}
&\partial_t^2 \zeta+2\Gamma_s \partial_t \zeta-s^2\partial_z^2 \zeta=2 \epsilon |b_0| \partial_t^2 a_1 ,\label{sm_corr_SU1} \\
&\partial_t a_1=-(\omega_p-\omega_x - \xi_0) a_2 -\Gamma_x a_1\:, \label{sm_corr_SU2} \\
&\partial_t a_2=(\omega_p-\omega_x - \xi_0) a_1 -\Gamma_x a_2-|b_0|\zeta\:. \label{sm_corr_SU3} 
\end{align}
Looking for the solution that depends on time and coordinate as $\e^{-\rmi \Omega t + \rmi k x }$ we obtain the solvability condition of the equations  (\ref{sm_corr_SU1})-(\ref{sm_corr_SU3}) in the form 
\begin{eqnarray}
\left(s^2 k^2-\Omega^2-2\rmi\Gamma_s\Omega\right)\left(\Delta^2+\Gamma_x^2-\Omega^2-2\rmi\Gamma_x\Omega\right) \nonumber \\
+ 2\Omega^2 \epsilon |b_0|^2 \Delta = 0 \,,\label{Eq_disp} 
\end{eqnarray}
where $\Delta=\omega_p-\omega_x - \xi_0$ is the detuning of the laser frequency from the exciton resonance.

In the limit of strong coupling, when $\Gamma_s$ and $\Gamma_x$ are small, the dispersion relation in the leading  approximation order is given by 
\begin{align}\label{Disp_strong_coupl_0} 
\Omega^2&= \frac{s^2k^2 +\Delta(\Delta-2\epsilon |b_0|^2 )}2
 \\ \nonumber
 &\pm \frac12 \sqrt{ \left[s^2k^2 -\Delta(\Delta+2\epsilon |b_0|^2 ) \right]^2-8\Delta^3\epsilon |b_0|^2} . 
\end{align}
For a positive laser detuning $\Delta>0$, the eigenfrequencies become complex in the vicinity of $k = \pm \sqrt{\Delta(\Delta+2\epsilon |b_0|^2 )}/s$. Moreover,  some eigenfrequencies have positive imaginary part, indicating that the corresponding eigenmodes grow with time. 

We now examine the effect of  finite losses on the instability. We assume that the losses and the pump are weak compared to the relevant sound frequencies, $\Gamma_x,\Gamma_s, \epsilon |b_0|^2 \ll |\Delta|, s|k|$. Then, it is natural to assume that  the most unstable mode will have the wave vector $k \approx  \pm |\Delta|/s$ and frequency $\Omega \approx |\Delta|$. In the vicinity of this point, Eq.~\eqref{Eq_disp} can be simplified to 
\begin{align}
(\Omega - s k + \rmi\Gamma_s) (\Omega - |\Delta| + \rmi\Gamma_x) = -\frac12\epsilon |b_0|^2 \Delta  \,.
\end{align}
The analysis yields that the instability persists if 
\[
\epsilon|b_0|^2  \Delta  > 2\Gamma_x\Gamma_s.  
\]
Specifically,  the acoustic waves with wave vectors in the region
\[
\left| |k| - \frac{\Delta}s\right| < \frac{\Gamma_s +\Gamma_x}s \sqrt{\frac{\epsilon |b_0|^2 \Delta}{2\Gamma_s\Gamma_x} -1} \,,
\]
are unstable.

\section{Optomechanical lasing regime}\label{sec:lasing}
\begin{figure}[t]
  \includegraphics[width=0.45\textwidth]{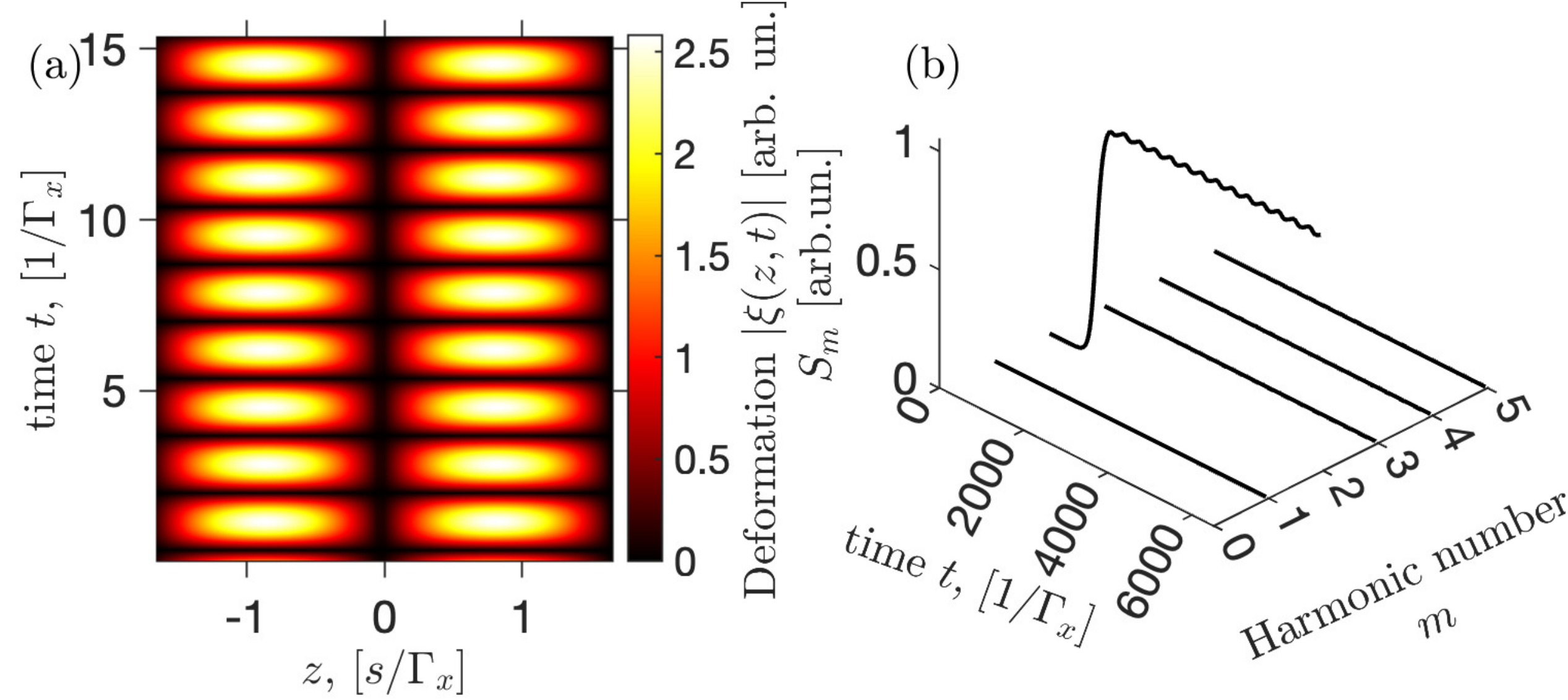}
  \caption{ (a) Stationary evolution of the deformation field $\xi(z,t)$. (b) Temporal evolution of the amplitudes of the first $5$ spatial harmonics. Calculation parameters are given in text.}
  \label{lasing_1}
\end{figure}

\begin{figure}[b]
  \includegraphics[width=0.47\textwidth]{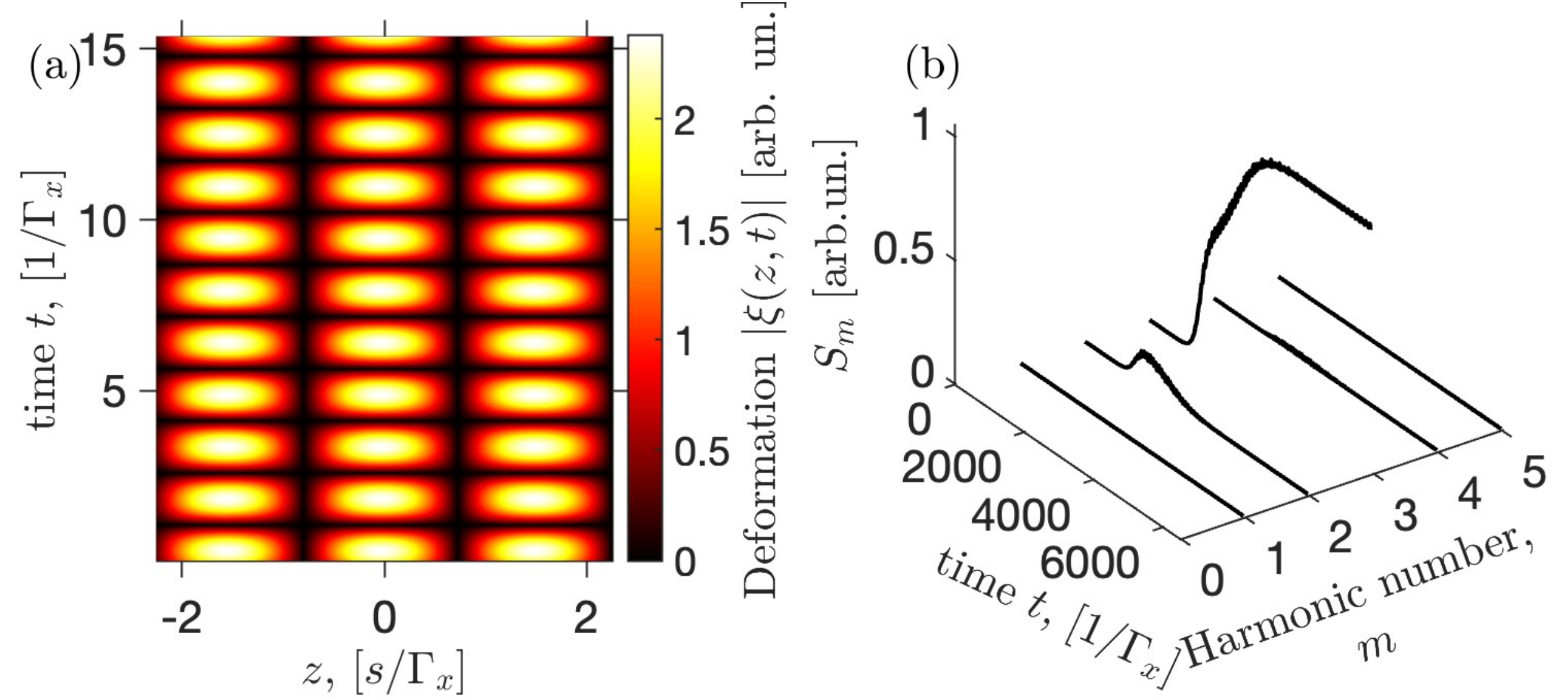}
  \caption{(Color online) (a) Stationary evolution of the deformation field $\xi(z,t)$. 
  (b) Temporal evolution of the amplitudes of the first $5$ spatial harmonics. Calculation parameters are given in text.}
  \label{lasing_2}
\end{figure}

We now  study numerically the system dynamics in the case when the ground state is unstable and the structure generates self-sustained oscillations. Figure~\ref{lasing_1} and Fig.~\ref{lasing_2} show the calculation results for  $\Delta/\Gamma_x=1.5$, $\Gamma_s/\Gamma_x=0.01$, $\epsilon G^2/\Gamma_x^3=0.09$,  and different system lengths $L$. We assume vanishing  stress at the structure edges, corresponding to  the boundary conditions $\xi(\pm L/2)  = 0$. The initial conditions are weak noise for $\xi$ and a stationary solution for $b_0$.

Figure~\ref{lasing_1} is calculated for a structure length $L=3.3\, s/\Gamma_x$. Panel (a) shows  the color plot of the deformation $\xi(z,t)$. Panel (b) presents the time evolution of the amplitudes of spatial spectral harmonics. We define them as $S_m=\sqrt{    (k_m/s)^2 |S_{\xi}(k_m)|^2     + |S_{\partial_t{\xi}}(k_m)|^2}$,  where $S_{\xi}$ and $S_{\partial_t{\xi}}$ are spatial spectra of $\xi$ and $\partial_t \xi$ correspondingly, and $k_m=\pi m/L$. One can see that the growth rate is positive only for one mode, $m=2$. 

For a wider system, several spatial harmonics can be located within the amplification range in the linear regime. The case when two harmonics are simultaneously amplified is  shown in Fig.~\ref{lasing_2}. The parameters are the same as for Fig.~\ref{lasing_1} but $L=4.5\, s/\Gamma_x$. The competition between the modes takes place and, depending on the initial condition, two different stationary states can form, where dominating is either the harmonic with $m=3$, as in Fig.~\ref{lasing_2}, or that with  $m=2$.

\begin{figure}
  \includegraphics[width=0.5\textwidth]{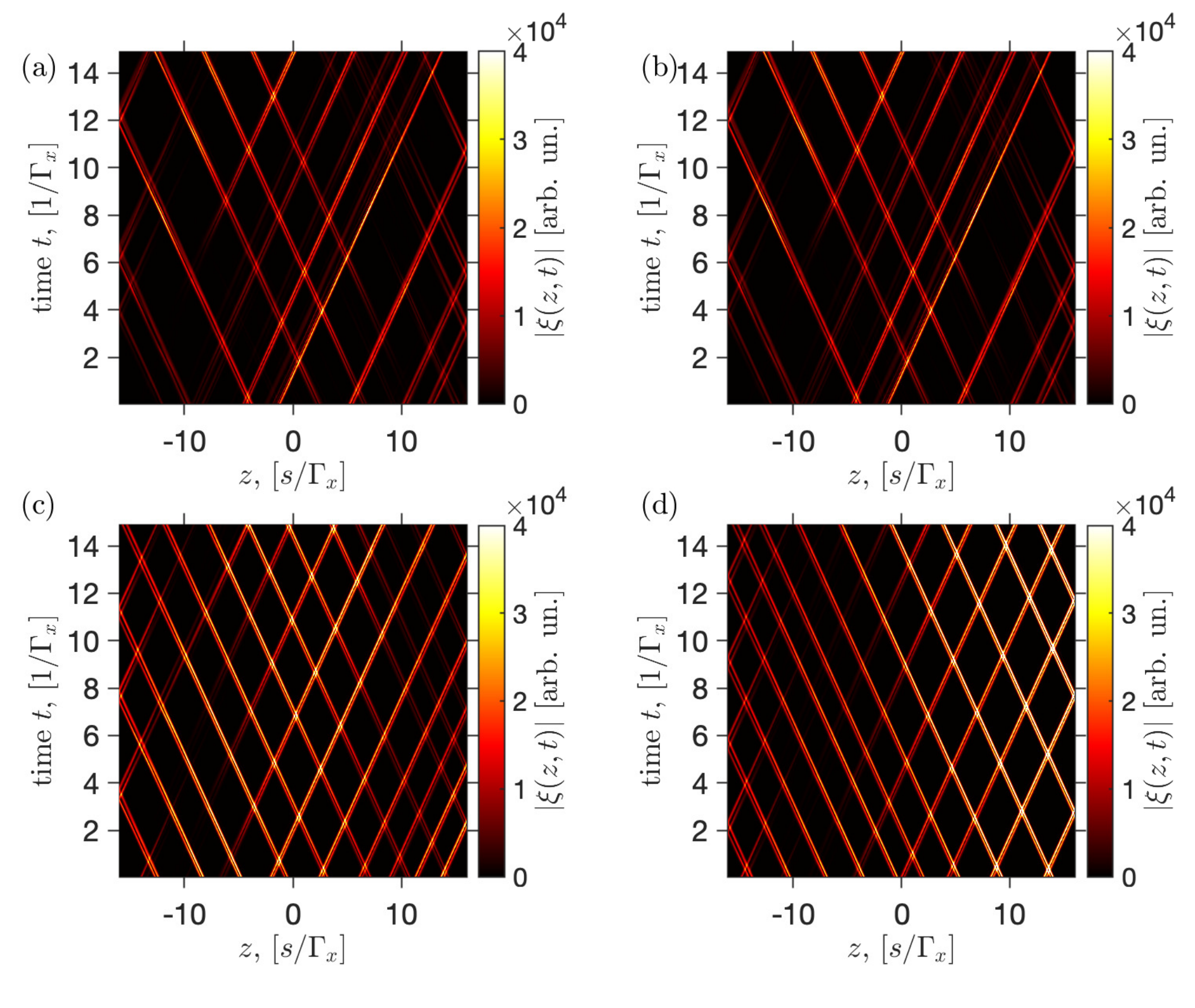}
  \caption{The stationary dynamics of the acoustic component  calculated for different acoustic impedances, determining the boundary conditions in Eq.~\eqref{eq:impedance}: $Z_r=0$ (a), $Z_r=0.25$ (b), $Z_r=4$ (c) and $Z_r=\infty$ (d). The other simulation parameters are $L=32s/\Gamma_x$ and 
  $\epsilon G^2/\Gamma_x^3=0.04$. 
  The initial conditions for the simulations were random noise in both acoustic and exciton components. In order to let the transitional processes decay the simulations results after a long delay $t_0=10 000/\Gamma_x$ are shown.    }
  \label{lasing_4}
\end{figure}

The dynamics becomes much richer in the case of a long system and high intensity of the pump. This ensures that a large number of spatial harmonics fall into the amplification range and get excited in the system simultaneously. At the nonlinear stage, the behaviour of the system is defined by a complex interplay of the interacting modes.
We performed numerical simulations for the pump $\epsilon G^2/\Gamma_x^3=4$ and the system length $L=32\,s/\Gamma_x$. The evolution of the acoustic energy distribution in the system is illustrated in Fig.~\ref{lasing_4}. Panels correspond to different boundary conditions imposed at the right edge, 
\begin{align}\label{eq:impedance}
\partial_t \xi_t (L/2) +s Z_r  \partial_z \xi (L/2) = 0 \,,
\end{align}
which describe the the situation when the area $z> L/2$ is filled with a material that has different acoustic impedance characterized by parameter $Z_r =0,$ 0.25, 4, $\infty$, for panels (a)--(d), respectively.  On the left edge we always suppose $\xi(-L/2) =0 $, i.e., $Z_l = 0$.

\begin{figure}
  \includegraphics[width=0.48\textwidth]{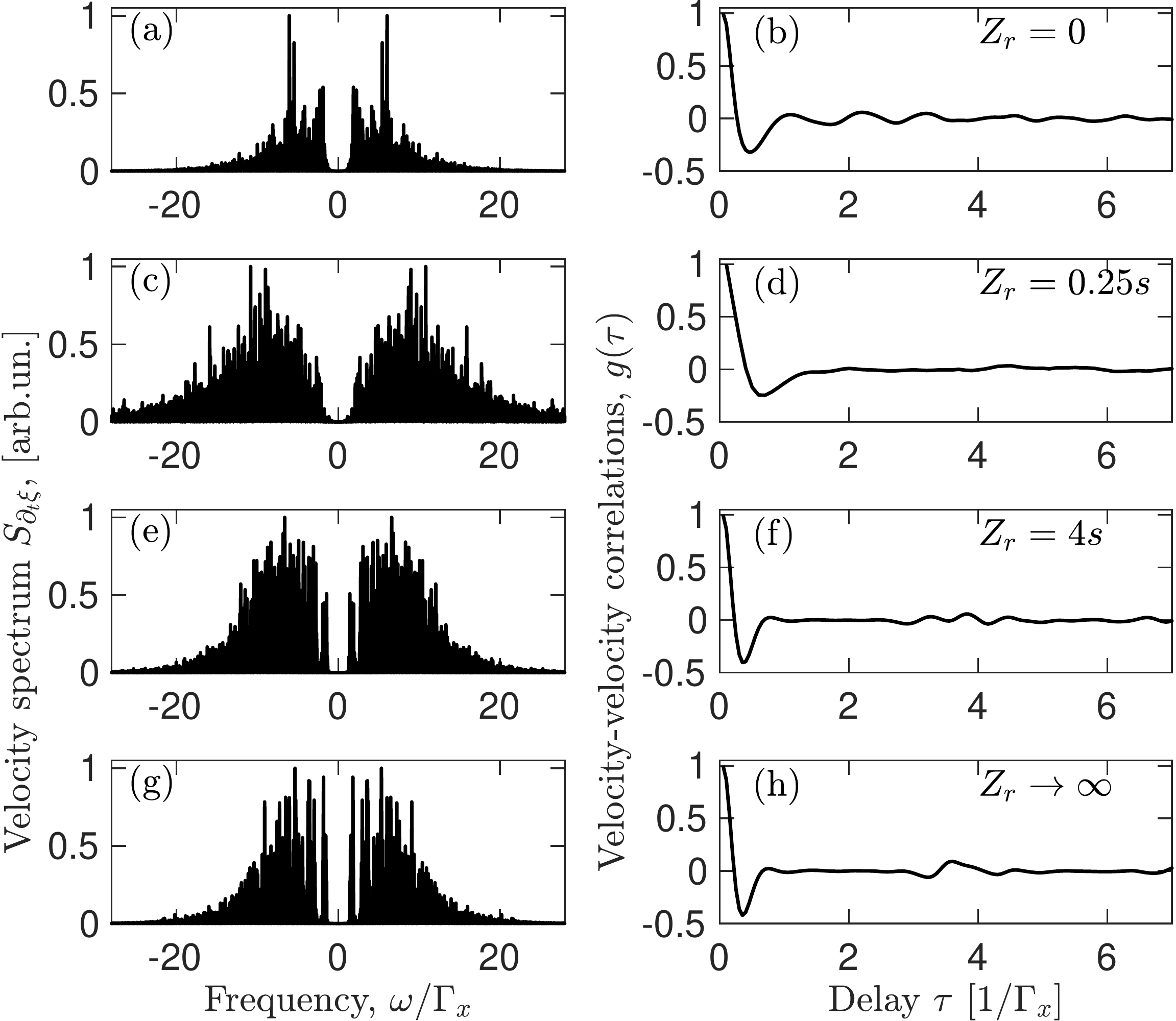}
  \caption{Spectra of the velocity (left panels) and corresponding velocity-velocity correlation functions $g(\tau)$ (right panels) calculated for different acoustic impedances at the right boundary $Z_r$.
For panels (a,b), the velocity was  measured at $z_0=0$ and for the rest at $z_0=L/2=16s/\Gamma_x$. The acoustic impedance $Z_r$ increases from top to bottom as indicated on graph.   The parameters are the same as for Fig.~\ref{lasing_4}.    }
  \label{lasing_5}
\end{figure}

One can see that the dynamics  varies significantly depending on the  impedance at the right edge. To further investigate this, we calculated the temporal spectra of the velocity $v(t)=\partial_t \xi(z_0,t)$, see Fig.~\ref{lasing_5} where the top row corresponds to  the impedances $Z_r=0$, the lower three rows correspond to  $Z_r=0.25s$, $Z_r=4s$ and $Z_r=\infty$ correspondingly. 
The velocity was calculated at the right end ($z_0=L/2$) for all panels excepting the top ones, Fig.~\ref{lasing_5}(a). The choice to measure the velocity at the right edge is made because this value defines the intensity of the radiation of the acoustic waves to the media contacting to the structure from the right. In the case when $Z_r=0$ the velocity at the right edge is exactly equal to zero, that is why we measured this value in the middle of the structure at $z_0=0$. 

One can see that all the velocity spectra  in Fig.~\ref{lasing_5}(a,c,e,g) are  quite broad but have different width and structure. The correlation functions $g(\tau)=\int v(t-\tau)v(t)\rmd t/\int v(t)^2 \rmd t $, shown in the right column of  Fig.~\ref{lasing_5}, decay rapidly which is in good accordance with the width of the spectra. The wide spectra and rapidly decaying correlation functions confirm that the dynamics of the acoustic field is very complex, probably chaotic. It is interesting to note that in the linear regime the absolute values of the acoustic wave reflection coefficients for $Z_r=0.25s$ and $Z_r=4s$ are the same, but in the nonlinear lasing regime the dynamics is different for these impedances. This can happen because the excitons affect the reflection of the acoustic mode at the edges of the system. 
\begin{figure}[b]
  \includegraphics[width=0.47\textwidth]{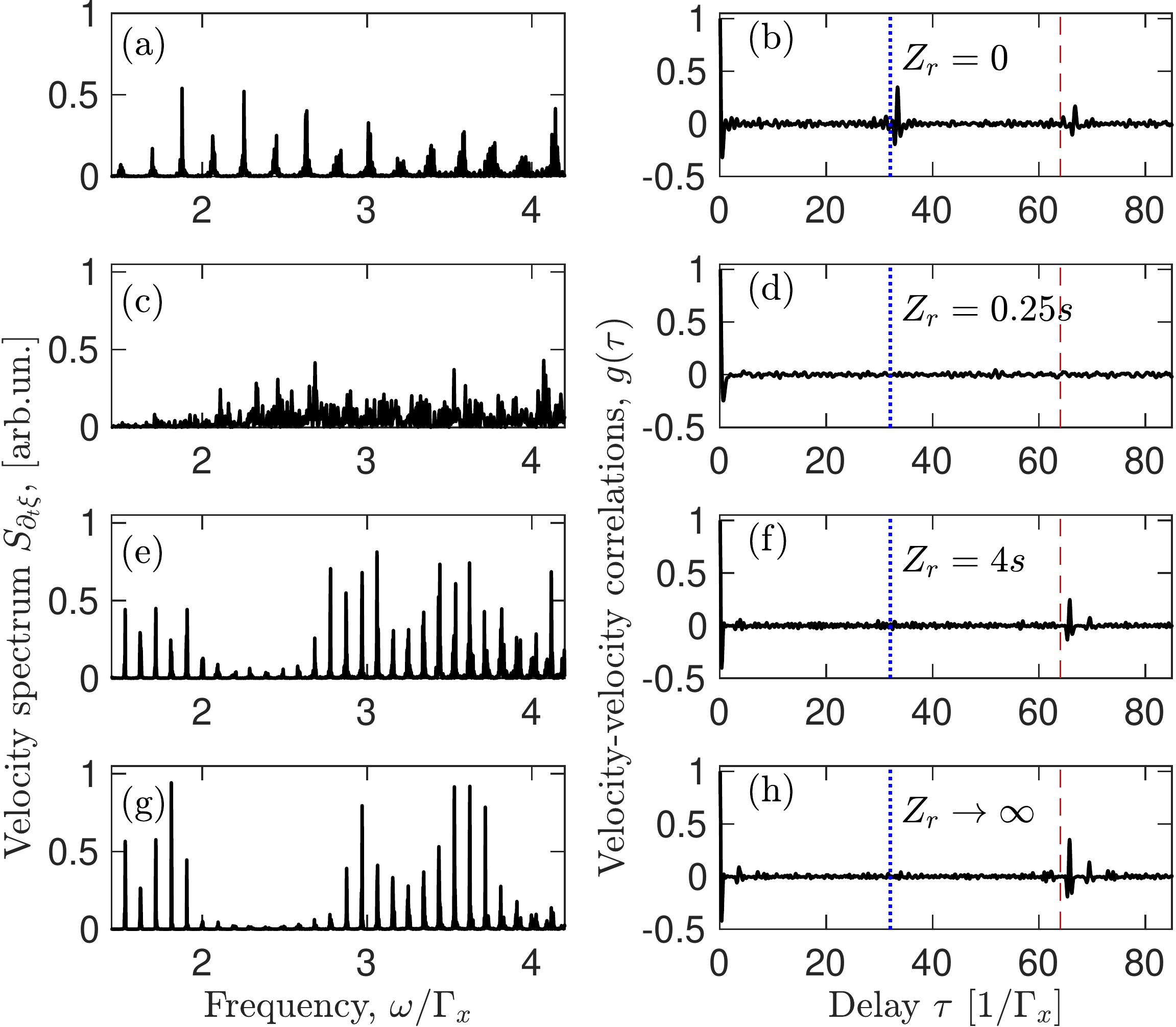}
  \caption{The same as Fig.~\ref{lasing_4} but a narrow part of the spectrum is shown with fine resolution. The correlation functions are shown for longer times. Vertical   lines in right panels indicate the times of acoustic wave trip from the left to the right and the roundtrip time.}
  \label{lasing_6}
\end{figure}  

It is instructive to look at the fine structure of the spectra and on the behaviour of the correlation functions at the times comparable to the travelling time of the acoustic waves through the system. This is illustrated in Fig.~\ref{lasing_6}. It is seen that in the case of $Z_r=0$ the correlation function has a sharp maximum at $\tau\approx 64/\Gamma_x$ which corresponds to the round trip of the acoustic wave in the system. Another maximum is seen at $\tau\approx 32/\Gamma_x$ which corresponds to the travelling time of the acoustic waves from the left to the right. For $Z_r=0.25s$ no maxima of the correlation function except the maximum at $\tau=0$ are seen. For $Z_r=4s$ and $Z_r=\infty$ the first maxima is situated at $\tau\approx 64/\Gamma_x$.   These features are also revealed in the fine structure of the spectra.  The separation between the neighbouring  modes corresponds to the above mentioned times: it  is $\Delta \omega \approx 0.2\Gamma_x$ in the top panel and $\Delta \omega = 0.1\Gamma_x$ in the two bottom panels. 

\section{Optomechanical domain walls}\label{sec:wall}
We now consider the situation when the pump is  {negatively} detuned from the excitonic resonance, {$\omega_p<\omega_x$}.
In this case the  background is stable and no optomechanical lasing occurs, but we show below that  optomechanical domain walls can form. 

We start with the case when  $\Gamma_s=0$ and rewrite the master equations~\eqref{eq:u2}-\eqref{eq:P2} in the reference frame moving with the velocity $v$,
\begin{eqnarray*}
&(\partial_t^2 -2v\partial_t\partial_\eta)(\xi - \epsilon |b|^2)=
(s^2-v^2)\partial_\eta^2\xi    +\epsilon v^2 \partial_\eta^2 |b|^2 ,\label{main_1_mov_RF} \\
&\partial_t b=v\partial_\eta b-\rmi(\omega_x-\omega_p) b -\Gamma_x b -\rmi b \xi -\rmi G\:, \label{main_2_movRF} 
\end{eqnarray*}
where $b(\eta,t) = P_1(z,t)$ with $\eta = z-vt$. 
The stationary solutions in the moving reference frame are governed by the coupled equations
\begin{eqnarray}
&\partial_\eta^2 \xi=\frac{\epsilon v^2}{v^2-s^2} \partial_\eta^2 |b|^2 ,\label{main_1_mov_RFst} \\
&v\partial_\eta b=\rmi(\omega_x-\omega_p) b + \Gamma_x b + \rmi b \xi + \rmi G\:.\nonumber
\end{eqnarray}
Equation (\ref{main_1_mov_RFst}) can be solved as $\xi=\epsilon v^2|b|^2/(v^2-s^2) +\xi_0$, where $\xi_0$ is a constant.
Using this we obtain the equation for the stationary distribution  of the exciton field $b$ 
\begin{eqnarray}
v\partial_\eta b=i\tilde \Omega b + \Gamma_x b + i \mu |b|^2 b + \rmi G \label{main_domain_wall} 
\end{eqnarray}
where $\tilde \Omega=\omega_x-\omega_p+\xi_0$ and $\mu=\epsilon v^2/(v^2-s^2)$.

Without a loss of generality we can fix that at $\eta=-\infty$ the state is characterized by $\xi=0$. This means that  $b(\eta=-\infty)=G/(\omega_p-\omega_x+\rmi\Gamma_x)$ and so $\xi_0=-\mu G^2/[\Gamma_x^2+(\omega_x-\omega_p)^2]$. This state we further refer as a basic state.

\begin{figure}
  \includegraphics[width=0.47\textwidth]{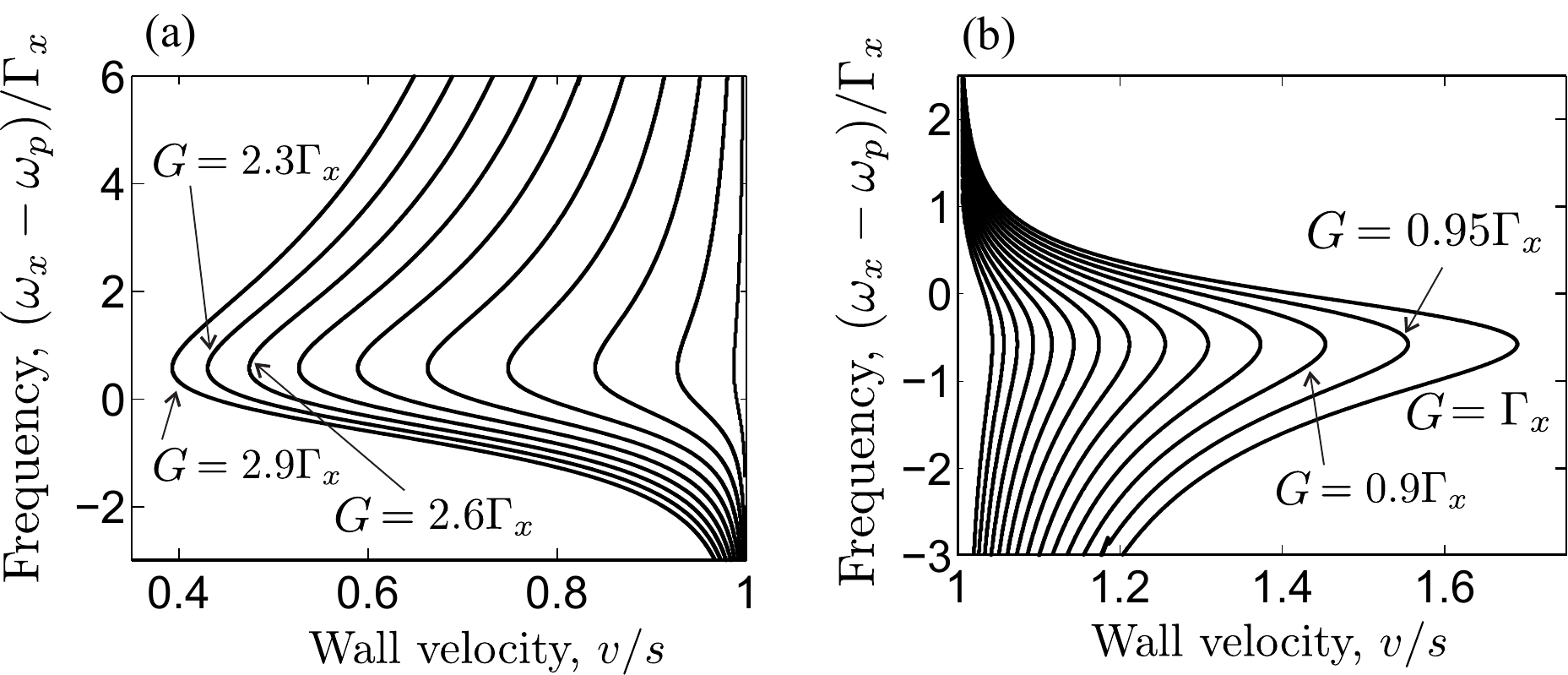}
  \caption{(Color online)  The $v-\omega_x$ diagram of  existence of subsonic and supersonic opto-acoustic domain walls for different amplitudes of the pump. In panel (a) the pump varies from $G=0.2\Gamma_x$ to $G=2.9\Gamma_x$ with the step $0.3\Gamma_x$, in panel (b) the pump varies from $G=\Gamma_x$ to $G=0.35\Gamma_x$ with the step $-0.05\Gamma_x$.  
}
  \label{existence_areas}
\end{figure}

\begin{figure}[b]
  \includegraphics[width=0.47\textwidth]{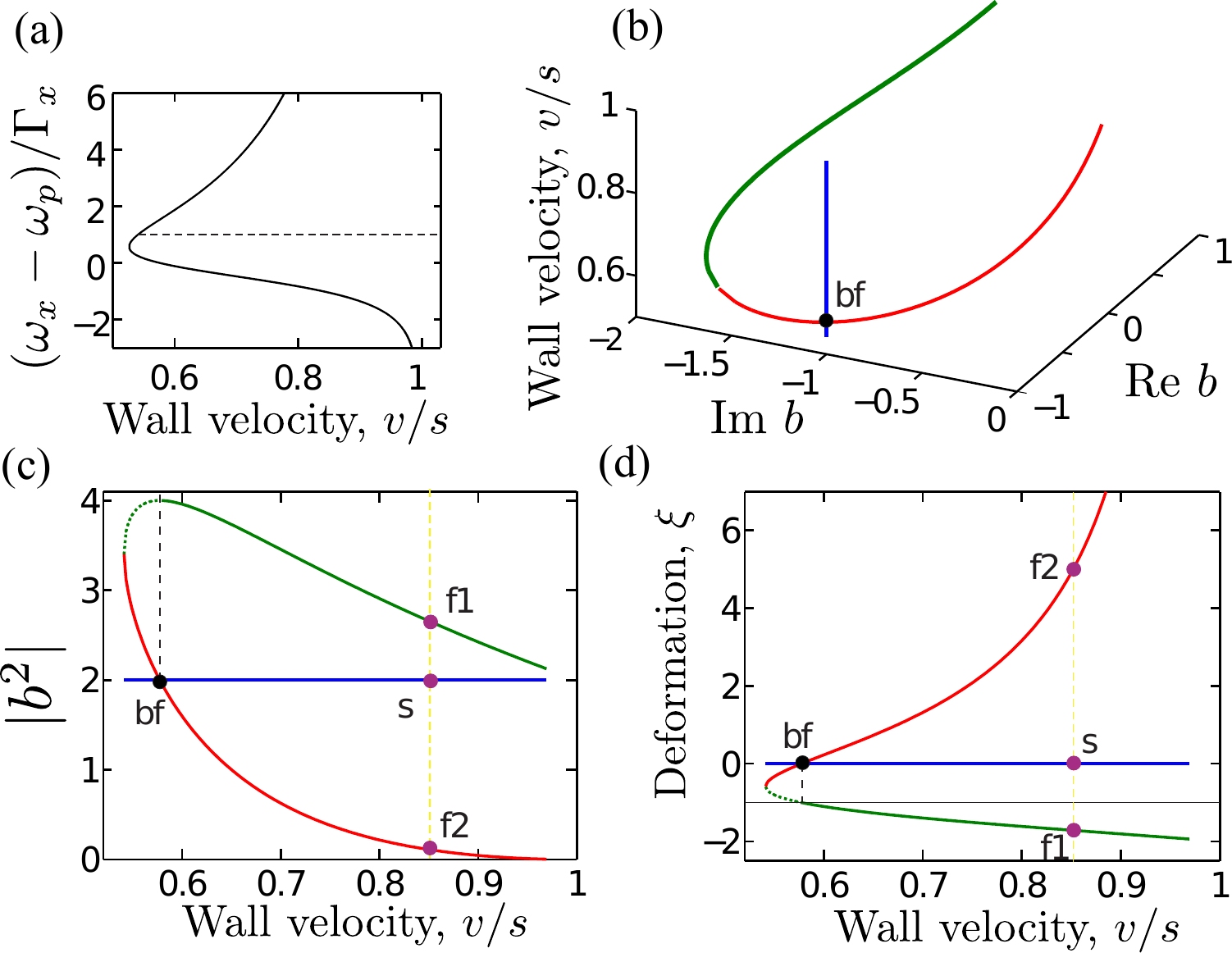}
  \caption{(Color online) Panel (a) shows the areas of multiple solutions for $G=2\Gamma_x$ and $v<s$. Panels (b) and (c) show the variation of the exciton component $b$ of the steady states with velocity $v$  [the path is shown by the dashed line in panel (a)] at $\omega_x-\omega_p=\Gamma_x$.   The blue lines correspond to the basic solution and the green and  red lines correspond to other two solutions. The collision of the basic state with another steady state is marked by ``bf''. The saddle and the two spiral steady states shown in the phase plane in Fig.~\ref{phase_plane1} are marked as ``s'', ``f1'' and ``f2'' correspondingly. The dashed line shows the steady state that does not have a connection to the basic state.}
  \label{steady_state_pos1}
\end{figure}

Let us find out if the basic state can be connected by a domain wall to another spatially uniform state. This is possible only if Eq.~(\ref{main_domain_wall}) has three spatially uniform solutions. It is straightforward to write the algebraic equation for the intensity of the spatially uniform states
\begin{eqnarray}
\mu^2 |b|^6 + 2\tilde \Omega \mu |b|^4 + (\tilde \Omega^2 +\Gamma_x^2)|b|^2-G^2=0. \label{st_states_ampl} 
\end{eqnarray}
Using the fact that one of the solutions is $|b_0|^2=G^2/[(\omega_x-\omega_p)^2 +\Gamma_x^2]$ we can represent the two other solutions of (\ref{st_states_ampl}) in a simple form
\begin{multline}
|b|^2=\frac{1}{2} \Bigl( \frac{G^2}{(\omega_x-\omega_p)^2+\Gamma_x^2} -\frac{2(\omega_x-\omega_p)}{\mu} \\\pm 
\sqrt{ \frac{G^4}{[(\omega_x-\omega_p)^2+\Gamma_x^2]^2} -\frac{4(\omega_x-\omega_p) G^2}{\mu[(\omega_x-\omega_p)^2+\Gamma_x^2]}-\frac{4\Gamma_x^2}{\mu^2} } \Bigr) . \label{another_2_sol} 
\end{multline}
Three solutions exist in the areas of parameters $\omega_x - v$ right of the curves in Fig.~\ref{existence_areas}(a) and left of the curves in Fig.~\ref{existence_areas}(b). 
Our analysis demonstrates that the supersonic domain walls can never connect dynamically stable backgrounds. That is why we focus our attention on the subsonic domain walls.
First of all we examine how the steady states changes when we vary the parameter $v$ for a fixed value of $\omega_x$ within the area of  multiple solutions, this path is shown by the dashed line in panel (a) of Fig.~\ref{steady_state_pos1}. The solution shown by the blue line corresponds to the basic solution $\xi=0$. The other two solutions are shown by the green and the red lines. Panels (b) and (c) show how the excitonic components of the steady states vary with the speed $v$. The analogous bifurcation diagram for the mechanical component $\xi$ is presented in panel (d). It can be seen that  at certain value of $v$ the basic steady state collides with the another steady state solution, the collision point is marked as ``bf'' in Fig.~\ref{steady_state_pos1}.
\begin{figure}
  \includegraphics[width=0.45\textwidth]{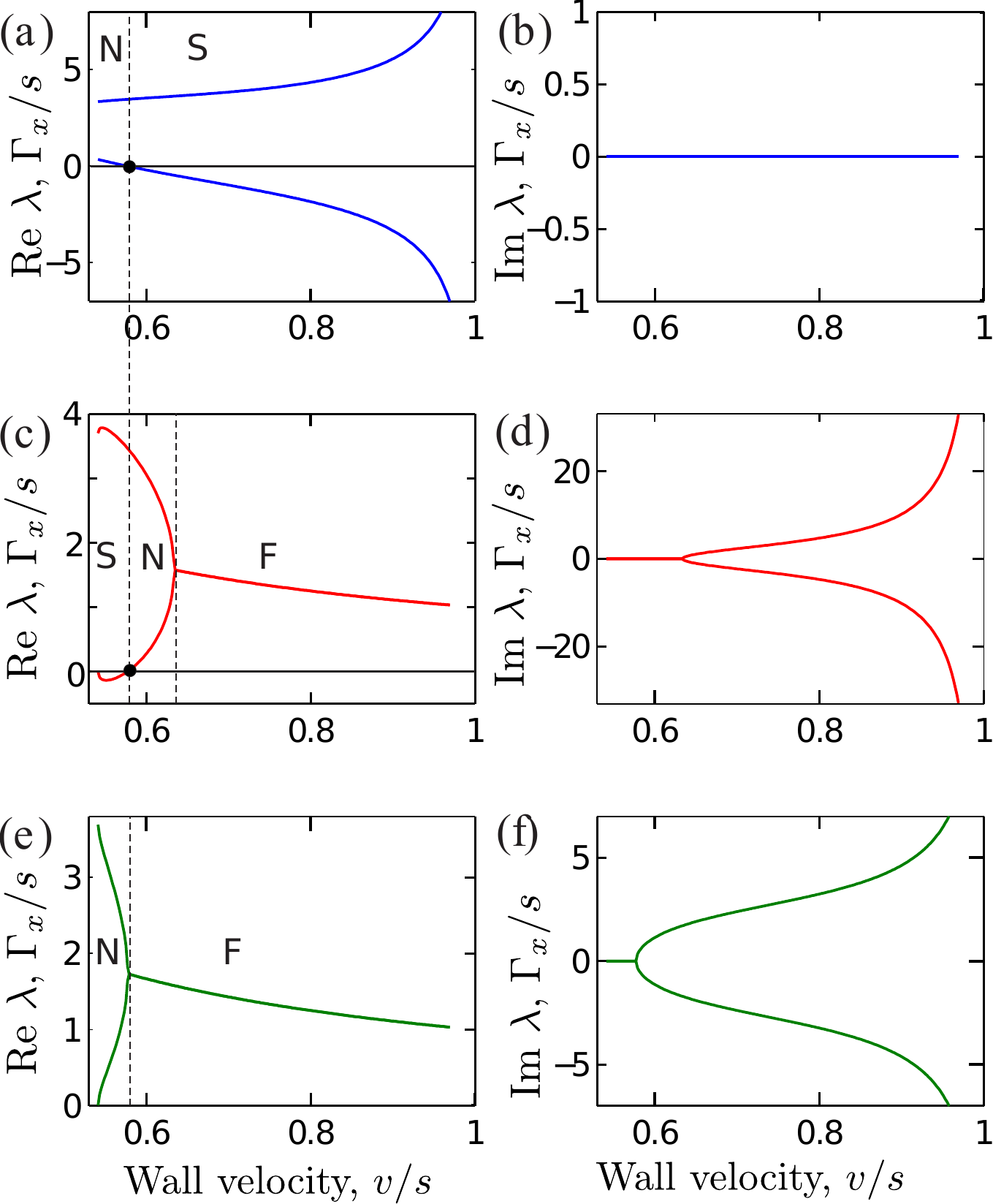}
  \caption{(Color online) The evolution of the real and imaginary parts of the eigenvalues governing the spatial relaxation  to the basic steady states is shown in panels (a) and (b) correspondingly. The same for the other two steady states is shown in panels (c)-(f). The types of the steady states are marked as ``s'' for saddles, ``n'' for knots and ``f'' for spirals (focuses). }
  \label{eig_val1}
\end{figure} 
\begin{figure}
  \includegraphics[width=0.5\textwidth]{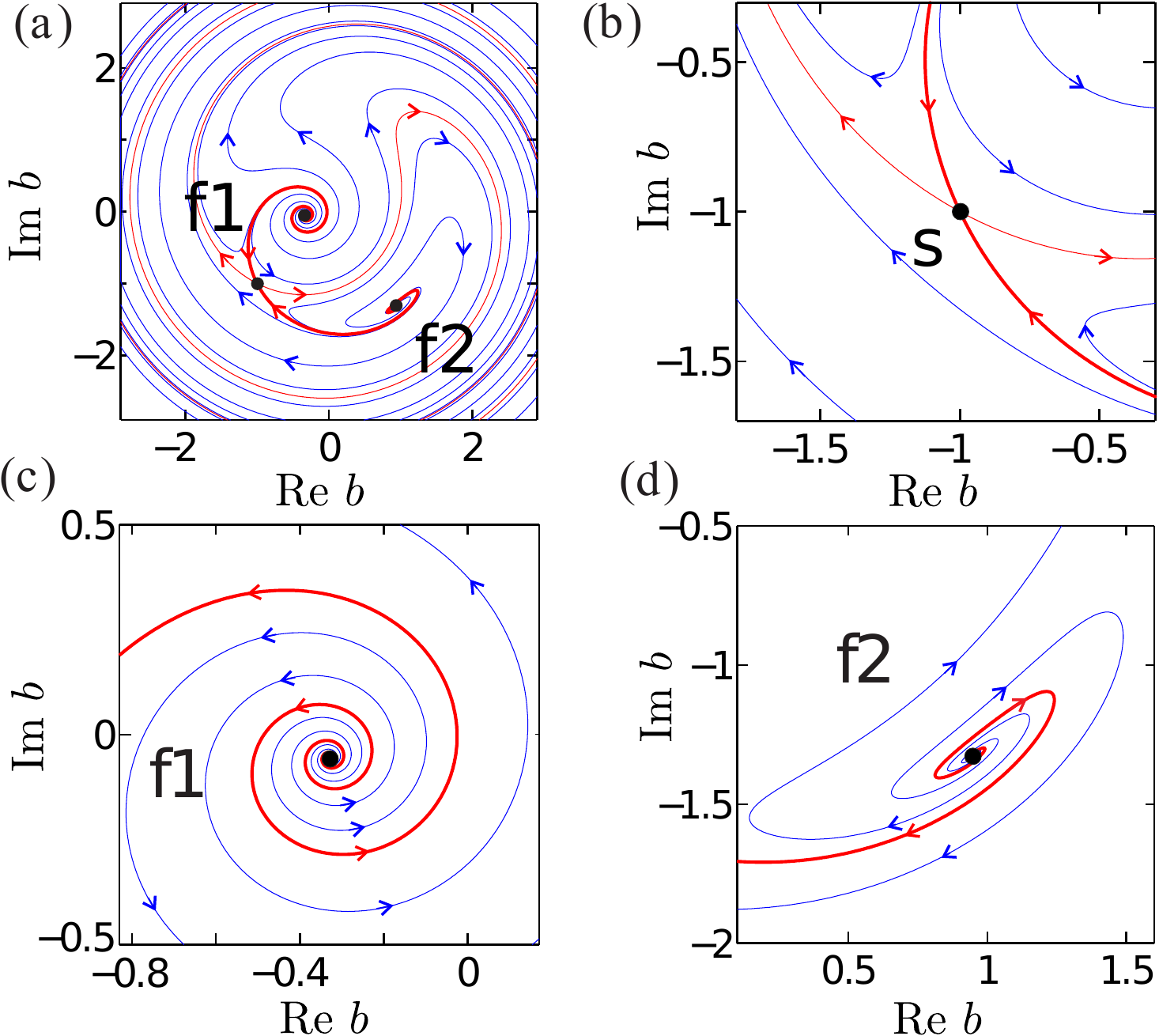}
  \caption{(Color online) Phase plane for $v=0.85s$ and $\omega_x=\omega_p+\Gamma_x$ is shown in panel (a). The parts of the phase plane in the vicinity of the steady states are shown in panels (b)-(d). 
  }
  \label{phase_plane1}
\end{figure} 
We now proceed to  classification of the steady states. This can be done by finding the eigenvalues of the linearized problem governing the spatial evolution of field $b$ in the vicinity of the equilibrium points. The real and imaginary parts of the corresponding eigenvalues are presented in Fig.~\ref{eig_val1}. 
The calculation shows that for the velocities close to $1$ the basic state is a saddle, the other two are focuses. The corresponding phase plane for $v=0.85s$ is shown in Fig.~\ref{phase_plane1}. It demonstrates that the basic state is a saddle indeed and it has heteroclinic connections to all other steady states, these heteroclinic connections are shown by thick black lines. This tells us that there exist two different domain walls moving with the velocity $v=0.85s$. These domain walls connect the basic state to the states with different mechanical $\xi$ and excitonic $b$ components.

\begin{figure}[t]
  \includegraphics[width=0.47\textwidth]{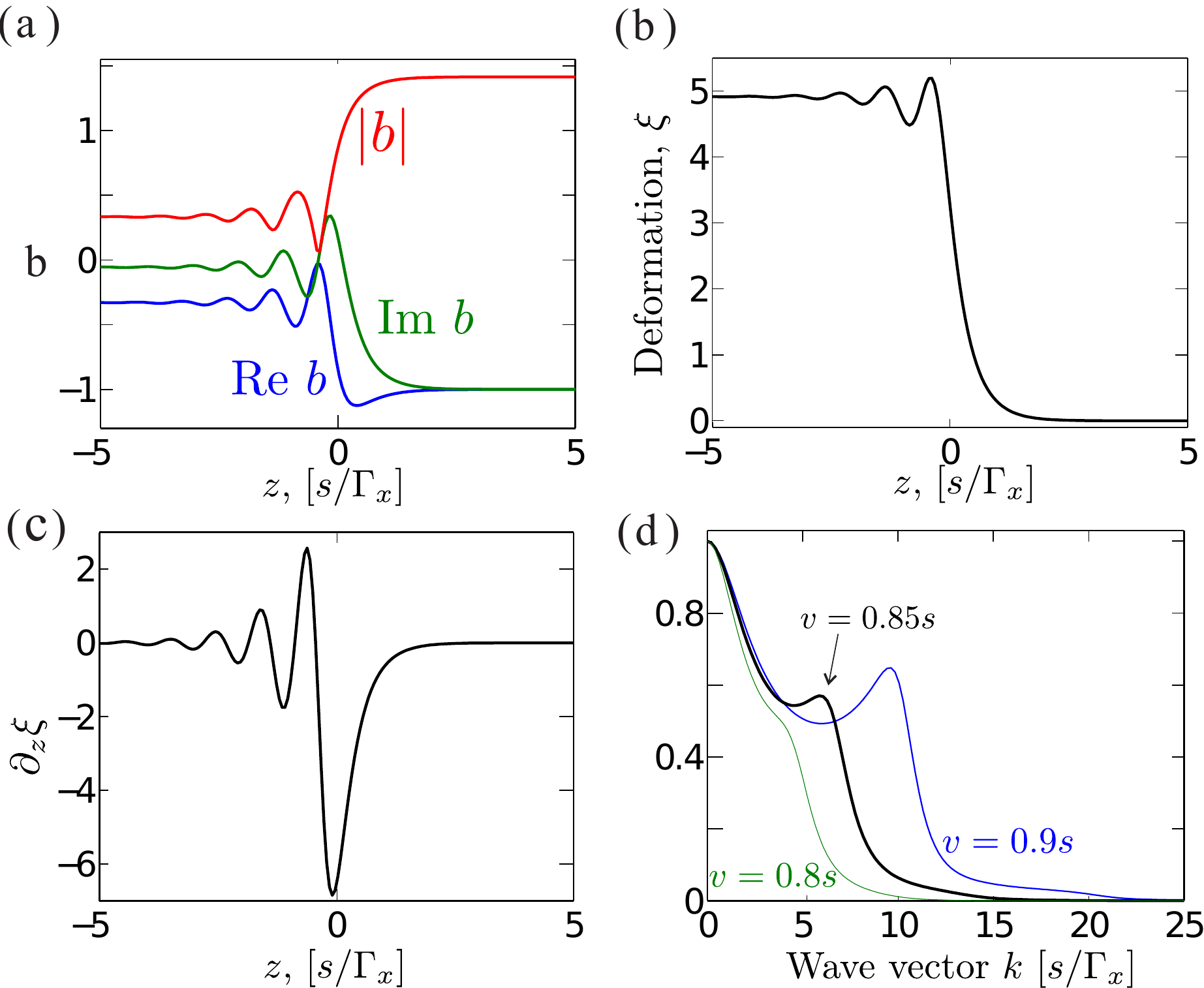}
  \caption{(Color online) The evolution of the excitonic field $b$ in the domain wall connecting the basic states ``s'' and the steady state ``f2''. Panel (b) and (c) show the normalized mechanical components $\xi$ and $\partial_z \xi$ in the domain wall. The normalized spatial spectrum of $\xi_z$ for $v=0.85s$ is shown in panel (d). The thinner blue and green lines show the spectra for the domain wall velocities $v=0.8s$ and $v=0.9s$. }
  \label{domain_wall}
\end{figure}

The spatial profile of the domain wall connecting the basic state to the state marked as ``f2'' is illustrated in Fig.~\ref{domain_wall}(a). Since one of the steady states is a spiral, the tail decaying to the steady state is oscillating in space. This is why  there exists an additional maximum in the spatial spectrum of the domain wall, see panel (b). This maximum becomes more pronounced for higher velocities when the oscillations become faster and the the decay rate gets slower, see panels (c) and (d) of Fig.~\ref{eig_val1}. For the lower velocities the relaxation becomes monotonic when the spiral transforms into knot. 

Our important result is  that at a threshold velocity the basic steady state collides with another steady state via a trans-critical bifurcation. At the bifurcation point the tails of the domain wall decay to the basic state algebraically because the corresponding eigenvalue turns to zero. It should be noted here that the basic state becomes a knot for the velocities below the threshold velocity of the bifurcation. This means that for these velocities the basic state can be connected to only one of the other two steady states. In the panels (c),(d) of Fig.~\ref{steady_state_pos1} the states not having a connection to the basic state are shown by the dashed line.

\begin{figure}[t]
  \includegraphics[width=0.5\textwidth]{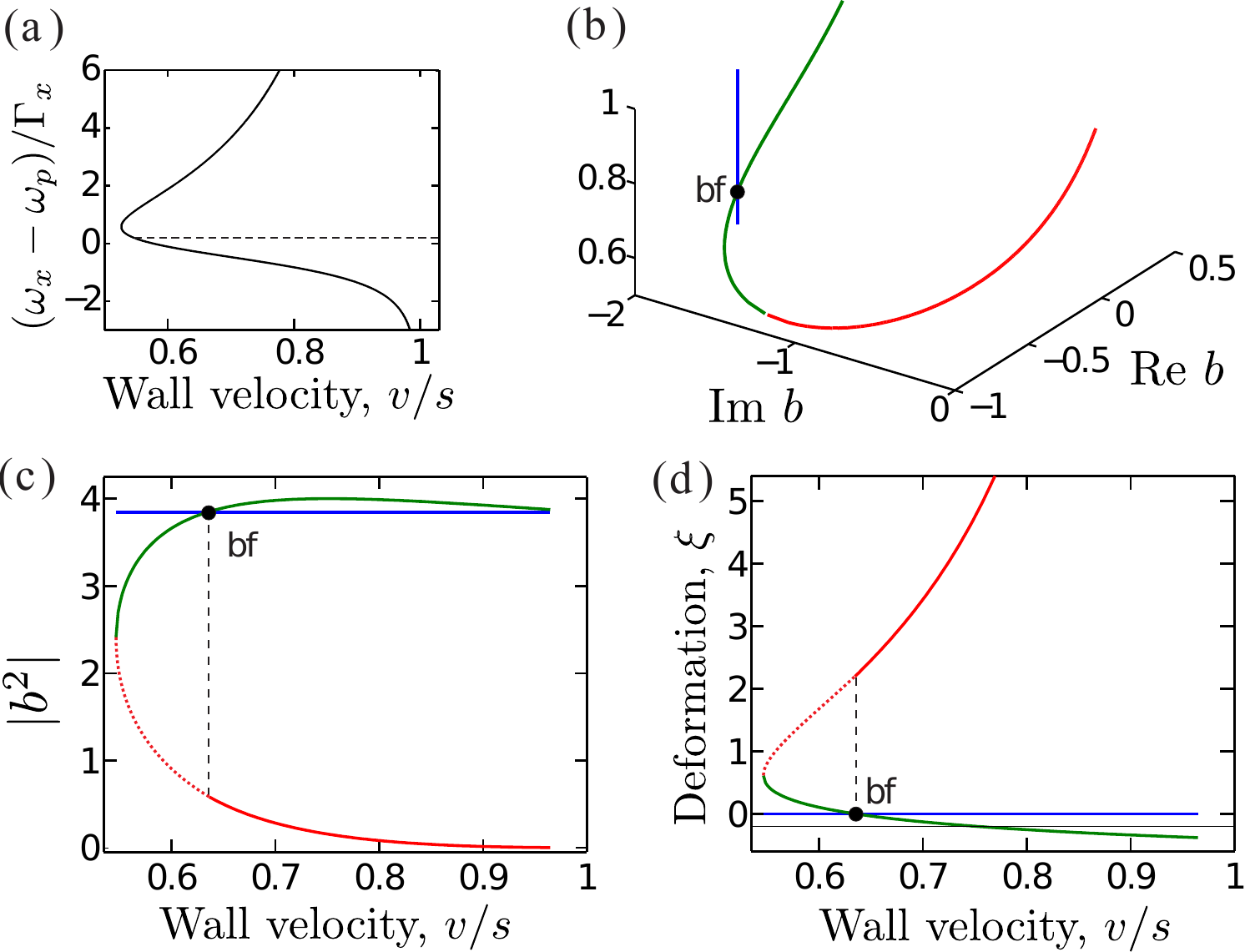}
  \caption{(Color online) The same as Fig.~\ref{steady_state_pos1} but for $\omega_x=\omega_p+0.2\Gamma_x$.
  }
  \label{steady_state_pos2}
\end{figure}

\begin{figure}[b!]
  \includegraphics[width=0.44\textwidth]{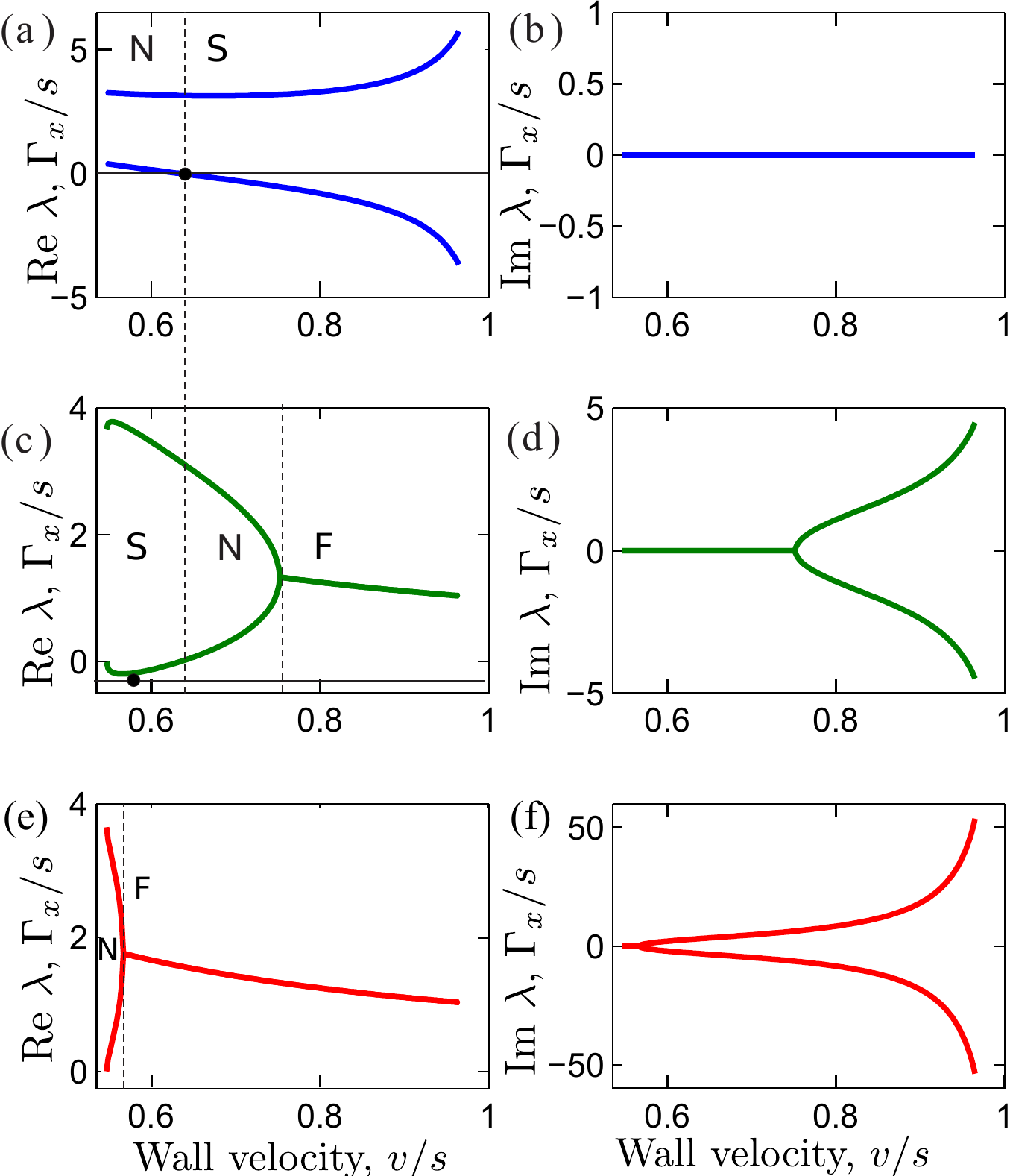}
  \caption{The same as Fig.~\ref{eig_val1} but for $\omega_x-\omega_p=0.2\Gamma_x$. 
  }
  \label{eig_val2}
\end{figure}

For the lower value of the frequency $\omega_x$ the bifurcation diagram looks quite differently, as shown in Fig.~\ref{steady_state_pos2}.  In this case the basic steady state collides with the other solution than for $\omega_x-\omega_p=\Gamma_x$. 
The dependences of the eigenvalues defining the relaxation of the domain walls to the backgrounds on the velocity $v$ are shown in Fig.~\ref{eig_val2}. At high velocities, the basic steady state is a saddle, the other two states are spirals. Then one of the spirals transforms to a knot. Then it no longer  collides with the basic state and a trans-critical bifurcation takes place: the basic state becomes a knot and the second state becomes a saddle. As in the previous case, after the bifurcation point the basic state can be connected to only one of other two states. At even smaller velocities, the second focus state transforms into a knot.

In order to be  experimentally observable   the domain walls have to be dynamically stable. The spectral analysis shows that this is the case provided that the backgrounds are stable. As it is discussed before, the spatially uniform states are stable if $(\omega_x-\omega_p)+\xi>0$. Thus, the stable domain walls are the connections between the states situated above the thin horizontal black lines in panel (d) of the Fig.~\ref{steady_state_pos1} and \ref{steady_state_pos2}. We have also examined how finite mechanical losses affect the domain wall propagation. It was found that the domain walls can form in the presence of mechanical losses provided that the losses are not too high.

The experiment on the propagation of the domain walls can be done as follows. After the formation of the spatially uniform state the left end of the system is shifted by some value $\Delta \xi$. This can be done in practice, for example, by application of a short optical laser pulse~\cite{Kalashnikova2020}. Then an optomechanical shock wave starts to propagate from the left end of the system to the right one. The formation of the domain walls was observed in direct numerical simulation of the master equations.
We have  checked numerically that the propagating waves are described well by the developed theory. Our numerical simulations also  indicate that the two domain walls form after the left edge of the system is shifted by $\Delta \xi_1$ at $t=t_1$ and then at $t=t_2>t_1$ by $\Delta\xi_2$ . In general the domain walls propagate with different velocities. If the second domain wall is faster than the first one, then at some moment they collide forming a new domain wall with its own velocity.

\section{Summary}\label{sec:summary}
To summarize, we have studied theoretically the strongly nonlinear regime of  optomechanical interaction between optically pumped excitons, localized in a semiconductor superlattice, and propagating acoustic phonons. We have examined different regimes of interaction depending on the detuning of the pumping frequency from the excitonic resonance.

When the structure is pumped above the excitonic resonance  it is in the optomechanical lasing regime. However, contrary to the usual situation of single-mode optomechanical laser~\cite{Kippenberg2014,Wu_2013,vyatkin2021optomechanical}, we reveal an intricate competition between  lasing acoustic modes that happens when the structure is long enough so that multiple spatial harmonics fall into the amplification range simultaneously and get excited. The wide spectra of the acoustic field and rapidly decaying time-dependent correlation functions indicate   quite complicated and probably chaotic lasing dynamics.  We also demonstrate that the spectra significantly depend on the boundary conditions for the acoustic wave at the edge of the structure.

In case when  the structure is pumped below the excitonic resonance, optomechanical lasing is not possible, but optomechanical domain walls can form instead. We perform a detailed  analysis of  steady states in the structure and   demonstrate, that only  subsonic domain walls can be stable and can  connect dynamically stable backgrounds. The dependence of the wall velocity and deformation amplitude on the pump frequency has been calculated.

We hope that our results provide useful insights in the rapidly developing field of resonant optomechanics and could be experimentally verified in the state-of-the-art structures. A natural extension of this work would be a study of the phonoritonic regime~\cite{Ivanov1982,Hanke1999,Poshakinskiy2017,Latini2021}, when not only excitons and acoustic waves, as in the current study, but three types of excitations, light, exciton and acoustic waves, experience a nonlinear interaction.

\section*{Acknowledgements}
This work has been funded by the Russian Science Foundation Grant No.~20-42-04405.

%

\end{document}